\documentclass[11pt]{article}

\usepackage[margin=1in]{geometry}
\usepackage{amsmath,amssymb}
\usepackage{graphicx}
\usepackage{booktabs}
\usepackage[numbers,sort&compress]{natbib}
\usepackage{xcolor}
\usepackage[font=small,labelfont=bf]{caption}
\usepackage[hidelinks]{hyperref}

\graphicspath{{figures/}}

\title{Phase-resolved dynamics of periodic lift-off and reattachment in an acoustically forced jet diffusion flame}
\author{Y. Sungtaek Ju\\[2pt]
  \small Department of Mechanical and Aerospace Engineering, University of California, Los Angeles,\\
  \small Los Angeles, CA 90095, USA \quad \texttt{sungtaek.ju@ucla.edu}}
\date{}

\begin{document}
\maketitle

\begin{abstract}
Laminar jet diffusion flames placed near a pressure node of a standing acoustic wave can switch from sustained oscillatory combustion (SOC) at the acoustic forcing frequency to periodic lift-off and reattachment (PLOR), a slow, large-amplitude cycle on which the forced response is superimposed. We analyze previously acquired high-speed imaging of a methane microjet flame, acoustically forced at 332~Hz with stepwise increasing amplitude, at jet Reynolds numbers of 20--100. Using proper orthogonal decomposition, separation of the slow and fast timescales, and a phase-resolved description of the slow cycle, we obtain four main results. (i) The two regimes are clearly distinct: the slow-variance fraction is at most 0.3\% in SOC and at least 67\% in PLOR. (ii) PLOR appears abruptly, the slow amplitude rising by a factor of 14--37 within a single 1.8~Pa amplitude step, with no measurable growth of slow fluctuations in the attached flame beforehand; the PLOR branch extrapolates to zero below the observed onset. These observations are consistent with a hysteretic onset rather than with the square-root growth expected at a supercritical Hopf bifurcation. (iii) The PLOR frequency, 15.9--16.7~Hz at onset for all Reynolds numbers, decreases to 10.4~Hz as the forcing increases because the cycle slows markedly at the phase of greatest lift-off: the flame spends up to 61\% of each period within one quarter of the cycle, while the excursion of the flame base grows to as much as 14.8 nozzle diameters. (iv) The forced response is modulated in amplitude and phase by the slow cycle, producing sidebands at the sum and difference frequencies in all leading modes rather than a mode-specific frequency-mixing signature. A compact phase-resolved model reproduces held-out data with a variance accounted for (VAF) of at least 0.90 when phase-locked to the observed cycle. We also identify potential sources of error specific to data of this kind: forcing harmonics aliased into the measurement band by the frame rate, bias of phase metrics referenced to the POD origin, filter transients in short records, and changes in camera framing between runs.
\end{abstract}


\section{Introduction}\label{sec:intro}

\subsection{Thermoacoustic coupling and diffusion flames}\label{sec:intro-coupling}

The interaction between acoustic waves and flames is a central problem in combustion science, with direct consequences for the stability of propulsion systems and industrial burners. Thermoacoustic instabilities arise from the coupling between unsteady heat release and acoustic pressure oscillations: when heat-release fluctuations occur sufficiently in phase with pressure fluctuations, acoustic energy is added to the system according to Rayleigh's criterion \cite{rayleigh1896}, and self-sustained oscillations of large amplitude can develop \cite{candel2002,lieuwen2012}. Their consequences include structural vibration, increased emissions, flame lift-off, and blow-off.

Both premixed and nonpremixed flames respond to acoustic velocity perturbations, but through different mechanisms. In premixed flames the response is governed largely by the kinematics of a propagating flame front, whereas in nonpremixed flames the heat release is controlled by the mixing of fuel and oxidizer, by the strain imposed on the reaction zone, and by the anchoring of the flame base \cite{candel2002,lieuwen2012,magina2019}. Forcing that perturbs the flame base can therefore change how the flame is stabilized, not merely modulate its heat release.

A well-controlled setting for such interactions is a fuel jet or droplet burning in a transverse standing acoustic wave. Droplet experiments showed that the response differs between pressure nodes and antinodes and that a burning droplet is deflected by the time-averaged acoustic radiation force acting on the hot, low-density region around it \cite{tanabe2000,tanabe2005,sevillaesparza2014}; under stronger forcing, periodic partial extinction and reignition were observed \cite{bennewitz2018}. For gaseous jets, \citet{sim2020} examined a methane microjet flame at $Re = 65$ near the pressure node of a 332~Hz standing wave. Phase-locked OH* chemiluminescence imaging showed that the flame was deflected away from the node and that, with increasing amplitude, its base lifted off and reattached in phase with the forcing. At higher amplitudes the phase-locked images became blurred and the Rayleigh index became negative, and time-resolved visible imaging at 1000~frames~s$^{-1}$ revealed an additional oscillation near 16--17~Hz, which decreased to about 14~Hz at the highest amplitude before blow-off. \citet{vargas2023} mapped the response of single- and multiport methane jet flames over a range of Reynolds numbers and forcing amplitudes. At low amplitude the flame remains attached to the burner and oscillates at the forcing frequency, a regime termed sustained oscillatory combustion (SOC). Above a threshold amplitude the flame base periodically detaches from and reattaches to the burner rim at a frequency far below that of the forcing, a regime termed periodic lift-off and reattachment (PLOR), and at still higher amplitude the flame blows off. The SOC--PLOR transition thus marks a change in the mode of flame stabilization; it is the subject of the present work.

\subsection{Flame lift-off in acoustically forced jets}\label{sec:intro-liftoff}

The transition of a jet flame from an attached to a lifted state is a classical problem, studied extensively in turbulent jets \cite{kalghatgi1984,pitts1989} and in laminar jets, where the lifted flame base takes the form of a tribrachial (triple) flame \cite{chung1991}. Acoustic forcing adds a time-dependent element: lifted nonpremixed flames subjected to longitudinal forcing can alternate between attached and lifted states \cite{baillot2002,baillot2007}, and buoyant laminar diffusion flames forced at low frequency exhibit synchronized vortex--flame interactions \cite{williams2007}. In laminar jets at low Reynolds number, which are steady without forcing, these phenomena can be examined without the complication of turbulent fluctuations.

In the near-pressure-node configuration of \citet{vargas2023}, the time-averaged radiation force deflects the flame away from the node, and during PLOR the deflected flame detaches periodically from the edge of the burner closest to the node before reattaching. Proper orthogonal decomposition (POD) of high-speed visible imaging showed that SOC is represented by a few modes, whereas in PLOR the fluctuation energy is distributed over many modes; the PLOR spectra contain a low-frequency peak (about 16~Hz in their example), the forcing frequency, and linear combinations of the two, indicating a quasi-periodic response. Related studies examined coaxial injectors near the pressure node \cite{vargas2025}, turbulent single and coaxial jets in a similar transverse-forcing configuration \cite{plascencia2024a,plascencia2024b}, and nine injector geometries forced at a pressure antinode \cite{hayrapetyan2025}. In the antinode configuration the perturbations are symmetric and PLOR appears predominantly as a period doubling at half the forcing frequency, whereas near the pressure node the lift-off frequency lies in the range 10--20~Hz and depends on the forcing amplitude \cite{hayrapetyan2025}. The two configurations therefore expose the flame to different forcing---mainly transverse velocity fluctuations and a mean radiation force near the node, mainly pressure fluctuations near the antinode---and produce different routes to lift-off.

\subsection{Bifurcation theory and data-driven modeling}\label{sec:intro-bifurcation}

Nonlinear dynamics and bifurcation theory provide a natural framework for describing such transitions \cite{guckenheimer1983,kuznetsov2004,strogatz2015}. A Hopf bifurcation, in which a fixed point loses stability to a limit cycle, underlies many thermoacoustic instabilities \cite{noiray2013,junipersujith2018}. Its two generic forms make distinct, testable predictions. At a supercritical Hopf bifurcation the oscillation amplitude grows continuously from zero, in proportion to the square root of the distance from onset, and noise-driven fluctuations of the stable state grow as onset is approached \cite{noiray2013}. At a subcritical bifurcation the amplitude jumps to a finite value, the system is bistable over a range of the control parameter, and the transition is hysteretic; describing it requires saturating nonlinearities of higher than cubic order \cite{subramanian2013,etikyala2017}. Distinguishing between the two from experimental data therefore requires examining the growth of the amplitude above onset and the behavior of the stable state below it and, ideally, measurements with the control parameter both increasing and decreasing.

When a self-excited oscillator is forced periodically, its response can lock in to the forcing or become quasi-periodic, with the two frequencies and their combinations present together \cite{lijuniper2013a,lijuniper2013b,kashinath2018,guan2019}; quasi-periodicity is also a recognized stage in routes to chaos in thermoacoustic systems \cite{kabiraj2012}. The forced jet flames considered here differ from these systems in an important respect: the unforced flame shows no detectable natural oscillation \cite{sim2020,vargas2023}, and the quasi-periodic state appears only when the forcing exceeds a threshold. The slow oscillation is therefore a property of the forced flame, arising from a bifurcation of its periodic forced response.

Data-driven reduced-order models offer a means of describing such dynamics directly from measurements. POD \cite{sirovich1987} reduces imaging data to a few modal coefficients, and sparse regression \cite{brunton2016} can identify governing equations for those coefficients. \citet{lettieri2025} applied sparse identification to under-sampled periodic data from these flames in the SOC regime; their analysis did not extend to PLOR. A description of the PLOR cycle itself, of how it changes with forcing amplitude and Reynolds number, and of the nature of its onset has not yet been reported.

\subsection{Scope and objectives}\label{sec:intro-scope}

This paper presents an analysis of high-speed imaging data for a single methane jet flame forced near the pressure node of a 332~Hz standing wave, at five Reynolds numbers and over the full range of forcing amplitudes from weak SOC to the last condition before blow-off. By separating the slow lift-off dynamics from the fast forced response, we address three objectives. First, we characterize the PLOR cycle in phase-resolved form and relate its phase to the geometry of the flame, in particular to the height of the flame base above the burner. Second, we determine the nature of the SOC--PLOR transition and the evolution of the cycle with forcing amplitude and Reynolds number. Third, we characterize how the forced response is modulated by the slow cycle. Because the forcing is sampled only about three times per period and each record lasts one second, we also examine the consequences of these sampling constraints for the analysis.

The paper is organized as follows. Section~\ref{sec:data} describes the data. Section~\ref{sec:methods} presents the analysis methods: the separation of timescales, the phase-resolved description of the slow cycle, the model of the forced response, and the diagnostics used to characterize the onset. Section~\ref{sec:results} presents the results, proceeding from the distinction between the two regimes and the nature of the onset to the structure of the PLOR cycle, its evolution with forcing amplitude, the modulation of the forced response, and the validation of the reduced description. Section~\ref{sec:discussion} discusses the interpretation and limitations of these results, and Section~\ref{sec:conclusions} presents the conclusions.

\section{Experimental data}\label{sec:data}

The data analyzed here were acquired in a previous experimental campaign with the apparatus and procedures described by \citet{vargas2023}. The experimental setup and conditions are briefly summarized here for the convenience of the reader. Methane was delivered via a stainless-steel tube of inner diameter $d_i = 0.457$~mm and outer diameter 0.794~mm into a closed, optically accessible cylindrical waveguide at atmospheric pressure. Loudspeakers at the two ends of the waveguide, 61~cm apart and driven out of phase, produced a standing wave at $f_a = 332$~Hz with a pressure node midway between them. The injector was placed 3~cm from the pressure node ($x/\lambda = +0.029$), where the flame experiences transverse acoustic velocity fluctuations and a time-averaged radiation force that deflects it away from the node. The jet Reynolds number $Re$, based on $d_i$ and the bulk exit velocity, took the values 20, 40, 60, 80, and 100.

For each Reynolds number, the forcing amplitude was increased stepwise and held constant at each level during recording. The amplitude is characterized by the maximum pressure perturbation in the waveguide, $p'_{\max}$, which was raised in 10~Pa increments from 100~Pa to the last level before blow-off (170, 190, 220, 240, and 260~Pa for $Re$ = 20, 40, 60, 80, and 100, respectively). The forcing is expressed here as the local pressure amplitude at the flame, $p' = 0.181\,p'_{\max}$, which is consistent with the paired values reported by \citet{vargas2023}; the 63 conditions thus span $p' = 18.1$--47.1~Pa in steps of approximately 1.8~Pa. No measurements were made with decreasing amplitude.

The visible (broadband) luminosity of the flame was recorded with a high-speed camera at 1000~frames~s$^{-1}$, with an exposure time approximately equal to the frame interval and a resolution of $512\times512$ pixels. Each condition is represented by a record of 1000 frames (1.0~s); these records were obtained in runs separate from those presented in \citet{vargas2023}. The images were preprocessed by pixel binning, minimum-intensity thresholding, and BM3D denoising \cite{dabov2007} and then decomposed by snapshot POD \cite{sirovich1987}; the present analysis uses the resulting mean field, the 20 leading spatial modes and their temporal coefficients, and the full eigenvalue spectrum of each record.

Image coordinates are expressed in nozzle diameters, with the nozzle exit at $r = z = 0$ and $z$ increasing downstream. The scale ($\approx$\,0.062\,$d_i$ per pixel) and the position of the nozzle within each frame were taken from calibrated plots produced with the original processing of these data; the nozzle position differs by up to about 3~$d_i$ between the series recorded at different Reynolds numbers, so all spatial comparisons between Reynolds numbers are made after registration to nozzle-centered coordinates. The calibration is consistent with the flame geometry: under the weakest forcing, the base of every flame lies 0.2--1.0~$d_i$ above the nozzle exit, its axis lies within 0.4~$d_i$ of the nozzle axis, and its length increases in proportion to the Reynolds number (Appendix~\ref{app:supp}, Figure~\ref{fig:coords}).

\section{Analysis}\label{sec:methods}

\subsection{Modal decomposition and separation of timescales}\label{sec:decomposition}

Each record is represented by its POD,
\begin{equation}\label{eq:pod}
I(\mathbf{x},t) \approx \bar I(\mathbf{x}) + \sum_{k=1}^{20} a_k(t)\,\psi_k(\mathbf{x}),
\end{equation}
where $I$ is the image intensity, $\bar I$ its temporal mean, $\psi_k$ the orthonormal spatial modes and $a_k$ the temporal coefficients. The eigenvalue associated with each mode is the intensity variance accounted for by that mode. The sign of each mode is arbitrary, so only sign-invariant quantities are interpreted; in particular, angles in the plane of two coefficients depend on the sign convention and are not assigned a physical meaning.

In PLOR the coefficients contain two well-separated timescales, the forcing period and the much longer lift-off cycle (Figures~\ref{fig:regimes} and~\ref{fig:twotime}). Each coefficient is therefore split into a slow and a fast part,
\begin{equation}\label{eq:split}
a_k(t) = a_k^{\mathrm{slow}}(t) + a_k^{\mathrm{fast}}(t) + a_k^{\mathrm{res}}(t),
\end{equation}
obtained with zero-phase, fourth-order Butterworth filters (low-pass at 50~Hz, high-pass at 100~Hz); the first and last 25 samples of each record are excluded from all statistics to avoid filter end effects. The residual band between 50 and 100~Hz contains at most 0.4\% of the variance in SOC and 2--8\% in PLOR, where it consists mainly of higher harmonics of the slow cycle.

The regime of each record is identified by the slow-variance fraction of the two leading modes,
\begin{equation}\label{eq:eta}
\eta = \frac{\operatorname{var}(a_1^{\mathrm{slow}}) + \operatorname{var}(a_2^{\mathrm{slow}})}{\operatorname{var}(a_1) + \operatorname{var}(a_2)},
\end{equation}
and the strength of the slow motion by
\begin{equation}\label{eq:A}
A^2 = \sum_{k=1}^{20} \operatorname{var}\!\left(a_k^{\mathrm{slow}}\right).
\end{equation}
Because the modes are orthonormal, $A^2$ is the slow-band variance of the image within the 20-mode subspace; unlike quantities built from individual modes, it does not depend on how the variance is distributed among the modes, which changes with Reynolds number and amplitude (Section~\ref{sec:res-spatial}). Uncertainties in $A^2$ and in the other record-level statistics are estimated from four contiguous blocks of each record, each containing about four slow cycles.

At 1000~frames~s$^{-1}$ the forcing is sampled about three times per period. Because the exposure time is close to the frame interval, harmonics of the forced response are attenuated but not eliminated, and they appear at the folded (aliased) frequencies $|n f_a - k f_r|$, where $f_r = 1000$~Hz is the frame rate. The second harmonic appears at $f_r - 2f_a \approx 336$~Hz, close to the forcing frequency itself; a 336~Hz component, consistent with this folding, was reported in dynamic mode decompositions of 1000~frames~s$^{-1}$ imaging of a similar flame \cite{sim2020}. The third harmonic appears at $3f_a - f_r = 3.985$~Hz, within the slow band, and a spectral line is observed at 3.983~Hz in the SOC records, growing with the forcing amplitude. Because it would otherwise be mistaken for slow flame motion, this line is removed from the SOC records by a least-squares fit of a sinusoid of known frequency. It is not removed from the PLOR records, because there it is negligible compared with the slow cycle, and a fit at this frequency over a one-second record would absorb part of the cycle itself.

\subsection{Phase-resolved description of the slow cycle}\label{sec:phase}

In PLOR the slow parts of the two leading coefficients trace a closed orbit, which is described in polar form,
\begin{equation}\label{eq:polar}
\left(a_1^{\mathrm{slow}} - c_1\right) + \mathrm{i}\left(a_2^{\mathrm{slow}} - c_2\right) = R\,\mathrm{e}^{\mathrm{i}\theta}.
\end{equation}
The reference point $(c_1, c_2)$ is the area centroid of the phase-averaged orbit. The POD origin, the natural alternative, is the time-averaged state, which is displaced toward the parts of the orbit where the system moves slowly (by 32\% of the mean radius in the median record and by up to 55\%); measuring the phase about it would understate how unevenly the orbit is traversed. The phase $\theta$ is unwrapped in time, and the frequency of the slow cycle, $f_s$, is obtained from a linear fit of $\theta(t)$, with an uncertainty estimated from the four blocks; direct cycle counting yields the same values within the uncertainties.

The shape of the cycle and the rate at which it is traversed are described by the radius and the rotation rate $\omega = \mathrm{d}\theta/\mathrm{d}t$ as functions of the phase,
\begin{equation}\label{eq:fourier}
\bar R(\theta) = r_0 + \sum_{n=1}^{N_R}\left(r_n^{c}\cos n\theta + r_n^{s}\sin n\theta\right),\qquad
\omega(\theta) = w_0 + \sum_{n=1}^{N_\omega}\left(w_n^{c}\cos n\theta + w_n^{s}\sin n\theta\right),
\end{equation}
fitted by least squares. The orders $N_R$ and $N_\omega$ are chosen by cross-validation in which each of the four blocks is predicted from the other three, taking the lowest order whose error is within one standard error of the minimum. Because the derivative in $\omega$ amplifies noise and a truncated series smooths sharp features, the rotation rate is also estimated without differentiation from the fraction of time spent at each phase. If $p(\theta)$ is the probability density of $\theta$ over the record and $T = 1/f_s$ the period,
\begin{equation}\label{eq:density}
\omega(\theta) = \frac{1}{T\,p(\theta)},
\end{equation}
where $p$ is estimated from a histogram with $5^\circ$ bins, smoothed with a Gaussian kernel with a standard deviation of $10^\circ$. Two measures quantify the nonuniformity: the ratio $\omega_{\min}/\langle\omega\rangle$ of the slowest to the mean rotation rate, and the dwell fraction, defined as the fraction of the period spent within $\pm45^\circ$ of the slowest phase, which equals 0.25 for uniform rotation. Both are computed from the density estimate~\eqref{eq:density}.

The slow parts of the higher modes are represented as functions of the same phase, $a_k^{\mathrm{slow}} \approx g_k(\theta)$ for $k = 3, 4$, using Fourier series of the form~\eqref{eq:fourier} with orders chosen in the same way. To relate the phase to the flame itself, the slow coefficients of all 20 modes are averaged within $10^\circ$ bins of $\theta$ and the corresponding phase-averaged images are reconstructed from~\eqref{eq:pod}. In each reconstruction, the flame is taken as the region in which the intensity exceeds the background by more than 30\% of the difference between the maximum of the mean field and the background; the height of its lowest point above the nozzle exit defines the height of the flame base, and its intensity-weighted centroid is also recorded.

\subsection{Forced response}\label{sec:forced}

The fast part of each coefficient is described as a response at the forcing frequency whose complex amplitude depends on the phase of the slow cycle, together with the folded second harmonic:
\begin{equation}\label{eq:forced}
a_k^{\mathrm{fast}}(t) \approx \operatorname{Re}\!\left[A_k\big(\theta(t)\big)\,\mathrm{e}^{2\pi\mathrm{i}f_a t}\right] + \operatorname{Re}\!\left[B_k\big(\theta(t)\big)\,\mathrm{e}^{2\pi\mathrm{i}(f_r-2f_a) t}\right],\qquad
A_k(\theta) = \sum_{m=-M}^{M}\alpha_{k,m}\,\mathrm{e}^{\mathrm{i}m\theta},
\end{equation}
with $B_k$ expanded in the same way. The coefficients $\alpha_{k,m}$ are obtained by linear least squares and the order $M$ by the cross-validation used for~\eqref{eq:fourier}; in SOC, where no slow cycle exists, $M = 0$. The forcing frequency is refined for each record by maximizing the variance captured by~\eqref{eq:forced}; it is 332.00~Hz in every record, with a total spread of 14~mHz. The form~\eqref{eq:forced} admits modulation of both the amplitude and the phase of the forced response: the terms $m = \pm1$ produce spectral lines at $f_a \pm f_s$, and higher $m$ produce lines at $f_a \pm m f_s$. The spatial pattern of the forced response at a given slow phase is
\begin{equation}\label{eq:pattern}
P(\mathbf{x},\theta) = \sum_{k=1}^{20} A_k(\theta)\,\psi_k(\mathbf{x}),
\end{equation}
whose modulus is independent of the sign convention of the modes.

\subsection{Diagnostics of the onset}\label{sec:onset-methods}

Three diagnostics address the predictions summarized in Section~\ref{sec:intro-bifurcation}. First, the ratio of the slow amplitude $A$ at the first PLOR condition to that at the last SOC condition measures the magnitude of the change across onset. Second, a supercritical onset implies $A^2 \propto p' - p'_c$ immediately above the critical amplitude $p'_c$. A straight line is therefore fitted to $A^2$ over the PLOR conditions of each Reynolds number and extrapolated to its zero, $p'_0$, with uncertainties estimated from a parametric bootstrap based on the block standard errors, from leave-one-out fits, and from a fit to the first three PLOR conditions only. If the onset were supercritical, $p'_0$ would lie between the last SOC and the first PLOR condition; $\Delta = p'_{\mathrm{SOC}} - p'_0 > 0$ indicates instead that the branch, continued downwards, lies below the observed onset. For comparison, a square-root law with its critical amplitude restricted to the observed onset interval is fitted to the same data. Third, a supercritical onset of a noise-driven system is preceded by growing fluctuations of the stable state. The power of the slow fluctuations in the band 8--30~Hz, which contains the PLOR frequency, is computed for every SOC record from its Hann-windowed spectrum; the window suppresses the end effects of the filters, which would otherwise grow with the forcing amplitude and mimic a precursor. Its trend with the forcing amplitude is tested with the Spearman rank correlation, and an extrapolation of its inverse to zero, which estimates the amplitude at which the attached state would lose linear stability, is reported only when the decline is robust.

\subsection{Validation}\label{sec:validation-methods}

The descriptions~\eqref{eq:fourier}--\eqref{eq:forced} are fitted to the first half of each PLOR record and evaluated on the second half. Two predictions are compared with the data: a phase-locked prediction, which uses the observed phase $\theta(t)$ and therefore tests the shape of the cycle and of the forced response, and a free-running prediction, which integrates $\mathrm{d}\theta/\mathrm{d}t = \omega(\theta)$ from the observed phase at the start of the test interval and therefore also tests the timing of the cycle. Agreement is measured by the variance accounted for,
\begin{equation}\label{eq:vaf}
\mathrm{VAF} = 1 - \frac{\operatorname{var}(y - \hat y)}{\operatorname{var}(y)},
\end{equation}
where $y$ is the measured and $\hat y$ the predicted signal. All estimators were also verified on synthetic records with the same format as the data and with prescribed properties (forcing frequency, slow frequency, nonuniform rotation, modulated forced response, folded harmonics, and noise); the prescribed values were recovered.

\section{Results}\label{sec:results}

\subsection{Sustained oscillation and periodic lift-off}\label{sec:res-regimes}

\begin{figure}[tbp]\centering\includegraphics[width=\textwidth]{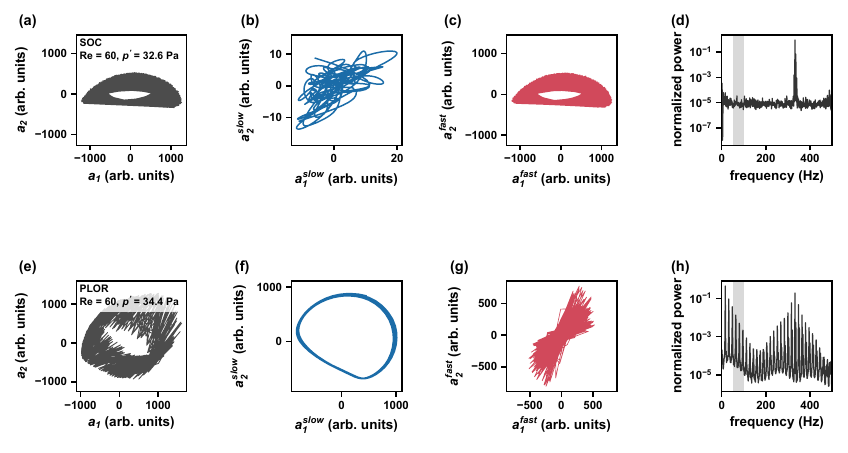}
\caption{SOC (top row, $Re = 60$, $p' = 32.6$~Pa) and PLOR (bottom row, $p' = 34.4$~Pa). (a,~e)~Coefficients of the two leading POD modes; (b,~f)~their slow parts ($<50$~Hz); (c,~g)~their fast parts ($>100$~Hz); (d,~h)~power spectrum summed over the 20 modes and normalized by its total, with the band 50--100~Hz shaded.}
\label{fig:regimes}\end{figure}

\begin{figure}[tbp]\centering\includegraphics[width=\textwidth]{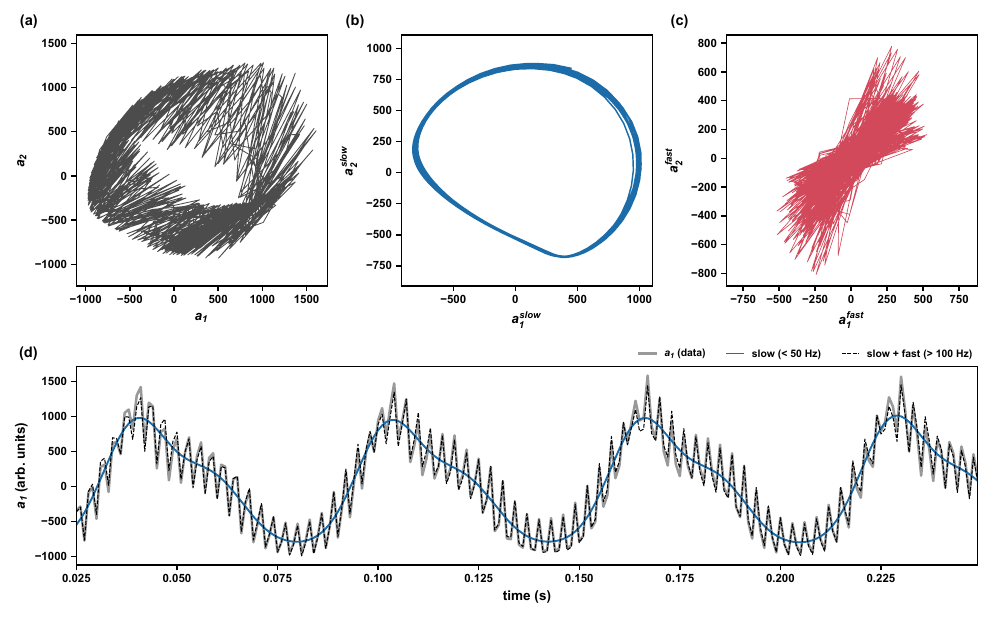}
\caption{Two-timescale structure of PLOR ($Re = 60$, $p' = 34.4$~Pa). (a)~Coefficients of the two leading modes; (b)~slow and (c)~fast parts; (d)~$a_1$ over 0.25~s with its slow part and the sum of its slow and fast parts.}
\label{fig:twotime}\end{figure}

Figure~\ref{fig:regimes} compares the last SOC and the first PLOR conditions at $Re = 60$. In SOC ($p' = 32.6$~Pa) the coefficients of the two leading modes trace a narrow band produced almost entirely by the response at the forcing frequency; their slow parts consist of small, irregular fluctuations, and the slow-variance fraction~\eqref{eq:eta} is 0.01\%. One amplitude step higher ($p' = 34.4$~Pa) the slow parts form a large closed orbit on which the forced response is superimposed, and the slow-variance fraction is 79\%. The spectra show the corresponding change: the forcing line dominates in SOC, whereas in PLOR a strong peak near 16~Hz and its harmonics appear, together with lines on either side of the forcing frequency. Figure~\ref{fig:twotime} shows that the slow and fast parts together reproduce the measured coefficients.

The distinction holds across all 63 conditions. The slow-variance fraction is at most 0.3\% in every SOC record and at least 67\% in every PLOR record, so the classification is insensitive to the choice of threshold, and it is unchanged for all 63 records when the low-pass cutoff is set to 40 or 60~Hz instead of 50~Hz. PLOR occurs at $Re$ = 40, 60, 80, and 100. At $Re = 20$ no PLOR condition precedes blow-off, in agreement with \citet{vargas2023}, and at $Re = 40$ PLOR is observed at a single condition before blow-off.

\subsection{Onset of PLOR}\label{sec:res-onset}

\begin{figure}[tbp]\centering\includegraphics[width=\textwidth]{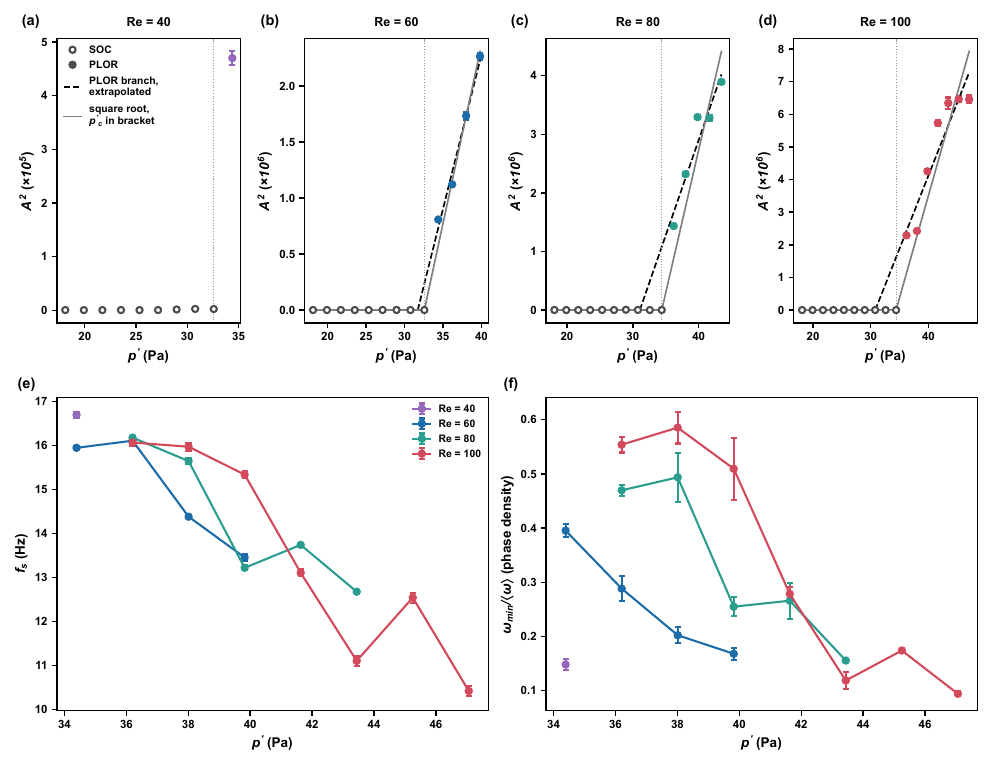}
\caption{Amplitude and frequency of the slow cycle. (a--d)~Slow amplitude $A^2$~\eqref{eq:A} against the local pressure amplitude for $Re$ = 40, 60, 80, and 100 (open symbols, SOC; filled symbols, PLOR; error bars, block standard errors); dashed lines, fitted to the PLOR conditions and extrapolated to zero; gray lines, square-root law with its onset restricted to the interval between the last SOC and the first PLOR condition; dotted lines, last SOC condition. (e)~Frequency of the slow cycle. (f)~Ratio of the slowest to the mean rotation rate from the phase density~\eqref{eq:density}.}
\label{fig:onset}\end{figure}

Table~\ref{tab:onset} and Figure~\ref{fig:onset}(a--d) characterize the onset. At every Reynolds number the slow amplitude $A$ increases by a factor of 14--37 between the last SOC and the first PLOR condition, that is, within a single amplitude step of 1.8~Pa. Above the onset, $A^2$ increases approximately linearly with $p'$, but the extrapolated zero of this branch lies below the last SOC condition: by $\Delta = 0.9$, 3.3, and 3.6~Pa at $Re$ = 60, 80, and 100, respectively. The estimates from the first three PLOR conditions alone, which are less affected by the leveling off of $A^2$ at the highest amplitudes, also lie below the last SOC condition (31.4, 33.4, and 32.5~Pa). At $Re = 60$ the result is marginal, the largest leave-one-out estimate coinciding with the last SOC condition, whereas at $Re$ = 80 and 100 all estimates lie below it. Correspondingly, a square-root law constrained to have its onset within the observed interval fits the PLOR data with a root-mean-square error 1.5--2.1 times larger than that of the unconstrained linear branch.

\begin{table}[tbp]\centering\small
\caption{Onset of PLOR. $A_{\mathrm{SOC}}$ and $A_{\mathrm{PLOR}}$ are the slow amplitudes~\eqref{eq:A} at the last SOC and the first PLOR condition; $p'_0$ is the zero of a straight line fitted to $A^2$ over the PLOR conditions, with its 95\% bootstrap interval, and over the first three PLOR conditions only; $\Delta$ is the amplitude of the last SOC condition minus $p'_0$ (all PLOR conditions). At $Re = 40$ a single PLOR condition precedes blow-off, and at $Re = 20$ none does.}
\label{tab:onset}
\begin{tabular}{rcccccc}
\toprule
$Re$ & last SOC $p'$ & first PLOR $p'$ & $A_{\mathrm{PLOR}}/A_{\mathrm{SOC}}$ & $p'_0$ (all levels) & $p'_0$ (first three) & $\Delta$ \\
 & (Pa) & (Pa) & & (Pa) & (Pa) & (Pa) \\
\midrule
20 & -- & -- & -- & -- & -- & -- \\
40 & 32.6 & 34.4 & 14 & -- & -- & -- \\
60 & 32.6 & 34.4 & 29 & 31.7 (31.6--31.9) & 31.4 & 0.9 \\
80 & 34.4 & 36.2 & 23 & 31.1 (30.7--31.3) & 33.4 & 3.3 \\
100 & 34.4 & 36.2 & 37 & 30.8 (30.4--31.1) & 32.5 & 3.6 \\
\bottomrule
\end{tabular}\end{table}

\begin{figure}[tbp]\centering\includegraphics[width=\textwidth]{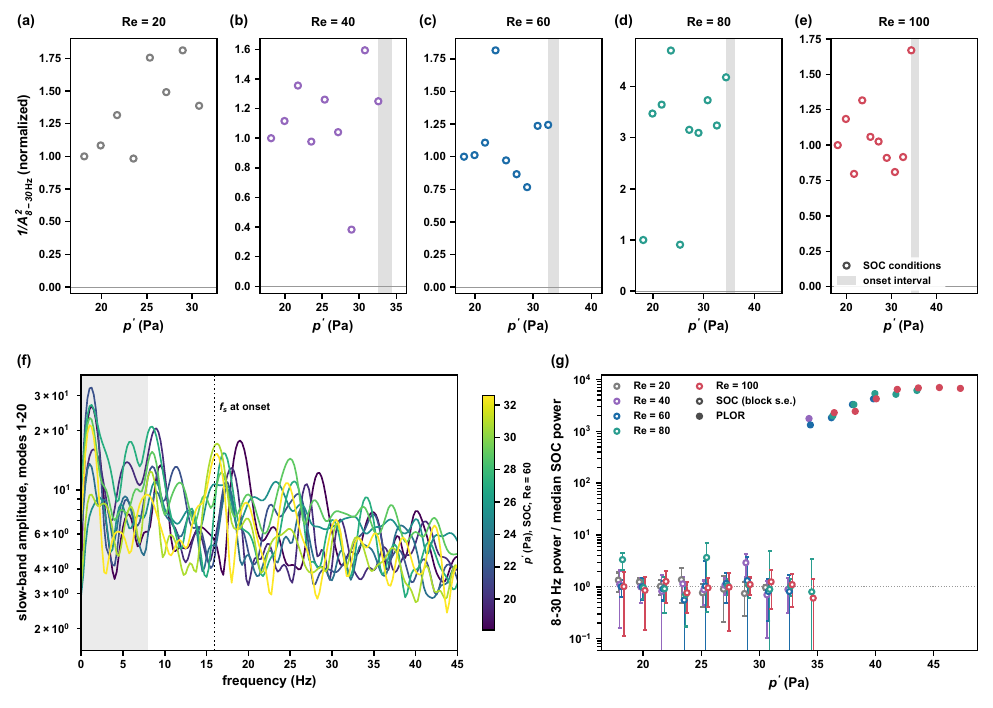}
\caption{Slow fluctuations of the attached flame. (a--e)~Inverse of the slow-band power in 8--30~Hz for the SOC conditions, normalized by its value at the lowest amplitude; gray bands, interval between the last SOC and the first PLOR condition. A linear-stability limit would appear as a systematic decline toward zero, which is not observed at any Reynolds number. (f)~Slow-band spectra of the SOC conditions at $Re = 60$ (the band below 8~Hz, shaded, is excluded from the analysis); dotted line, frequency of the slow cycle at onset. (g)~Power in 8--30~Hz for every condition, normalized by the median over the SOC conditions of each Reynolds number (open symbols, SOC with block standard errors; filled symbols, PLOR).}
\label{fig:precursors}\end{figure}

The attached flame shows no sign of the approaching transition (Figure~\ref{fig:precursors}). The power of its slow fluctuations in the band 8--30~Hz, which contains the PLOR frequency, does not increase with the forcing amplitude at any Reynolds number: the Spearman correlation with $p'$ ranges from $-$0.69 to $-$0.02 and is not significant in any case ($p \geq 0.06$). The spectra of the SOC records show no peak that grows near the eventual PLOR frequency, and the inverse band power shows no robust decline toward zero; consequently, no amplitude of linear instability can be extrapolated. At onset, the band power rises by a factor of approximately 1300--2300 within one amplitude step.

\subsection{Structure of the PLOR cycle}\label{sec:res-cycle}

\begin{figure}[tbp]\centering\includegraphics[width=\textwidth]{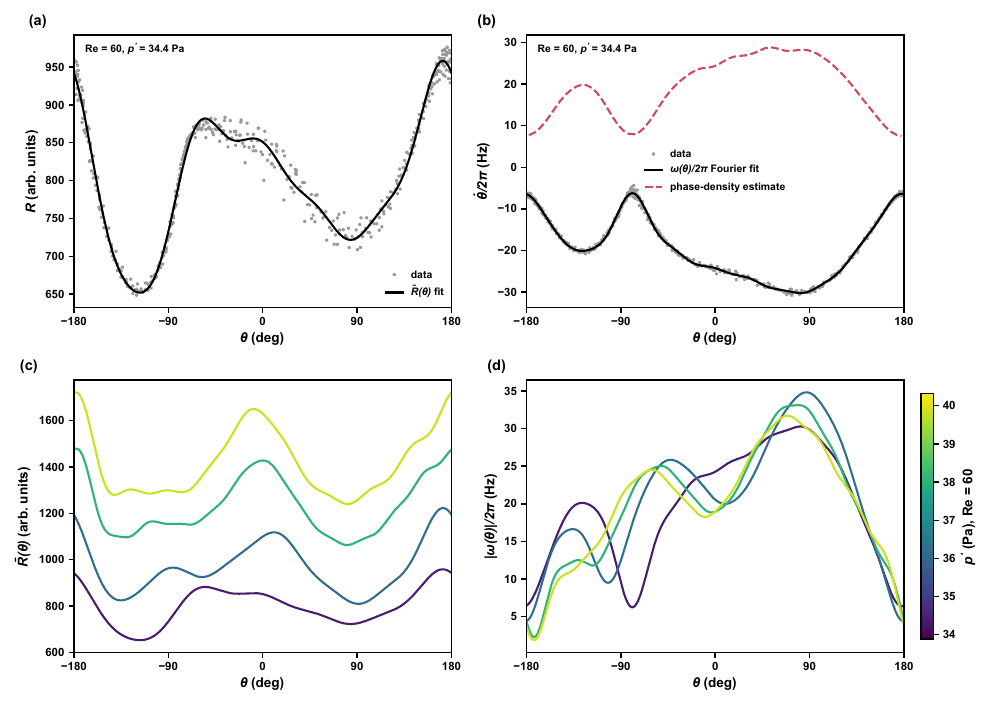}
\caption{Phase-resolved description of the slow cycle. (a)~Radius $R$ against phase with the fit $\bar R(\theta)$ and (b)~rotation rate $\dot\theta/2\pi$ with the Fourier fit and the phase-density estimate~\eqref{eq:density}, for $Re = 60$, $p' = 34.4$~Pa. (c)~$\bar R(\theta)$ and (d)~$|\omega(\theta)|/2\pi$ for all PLOR conditions at $Re = 60$, colored by $p'$.}
\label{fig:slowcycle}\end{figure}

Figure~\ref{fig:slowcycle}(a,~b) describes the slow cycle at $Re = 60$, $p' = 34.4$~Pa ($f_s = 15.94$~Hz). The radius varies with phase, so the orbit is not circular, and it is traversed at a strongly nonuniform rate: the rotation rate falls to 0.40 of its mean near $\theta = 178^\circ$, and the flame spends 0.36 of the period in the quarter-cycle centered on that phase. The Fourier descriptions~\eqref{eq:fourier} require orders of 6 and 10 for $\bar R$ and $\omega$, respectively, and account for 0.98 and 0.996 of the variance in the held-out blocks; the rotation rate obtained from the phase density agrees with the Fourier description.

The phase-averaged reconstructions (Figure~\ref{fig:physical} and Figure~\ref{fig:S1} in Appendix~\ref{app:supp}) relate this description to the flame. The flame is deflected away from the pressure node, to positive $r$, and its centroid moves by 1.7~$d_i$ in $r$ over the cycle. The flame base is lowest near $\theta = -55^\circ$, where it lies slightly below the nozzle exit plane, as does the base of the unforced flame on this burner \cite{sim2020}, and highest near $\theta = 165^\circ$, a lift of 1.6~$d_i$; the slow portion of the cycle therefore coincides with the phase at which the flame base is most lifted. Lift-off and reattachment thus proceed comparatively rapidly, whereas the flame spends a disproportionate part of each cycle in its lifted configuration.

\begin{figure}[tbp]\centering\includegraphics[width=\textwidth]{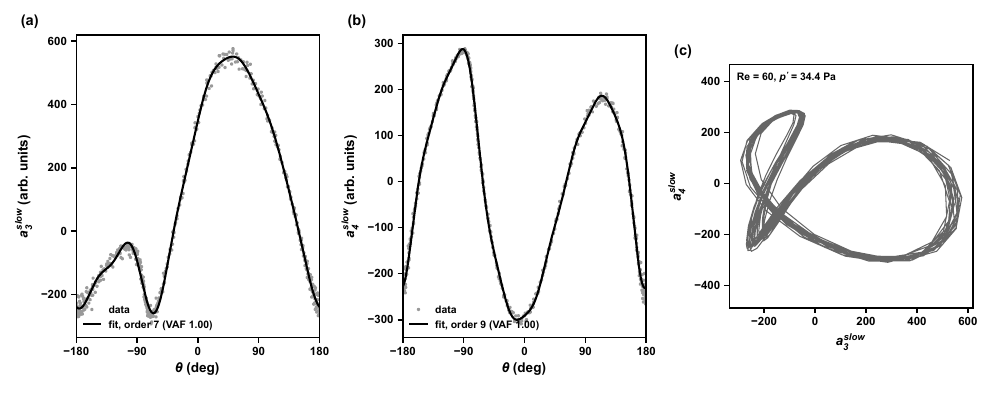}
\caption{Slow parts of modes 3 and 4 against the phase of the slow cycle at $Re = 60$, $p' = 34.4$~Pa: (a)~$a_3^{\mathrm{slow}}$ and (b)~$a_4^{\mathrm{slow}}$ with their Fourier fits; (c)~projection on the $(a_3^{\mathrm{slow}}, a_4^{\mathrm{slow}})$ plane.}
\label{fig:slaving}\end{figure}

The slow parts of modes 3 and 4 are single-valued functions of the phase (Figure~\ref{fig:slaving}), which account for 0.996 and 0.997 of their variance at this condition and between 0.93 and 0.999 over all PLOR conditions. At the first PLOR conditions for $Re$ = 60 and 80, mode~3 varies mainly at the fundamental of the slow cycle and mode~4 at its second harmonic; at higher amplitudes, and at all amplitudes for $Re = 100$, both are dominated by the second harmonic. The slow dynamics of the four leading modes are thus organized by a single phase variable.

\begin{figure}[tbp]\centering\includegraphics[width=\textwidth]{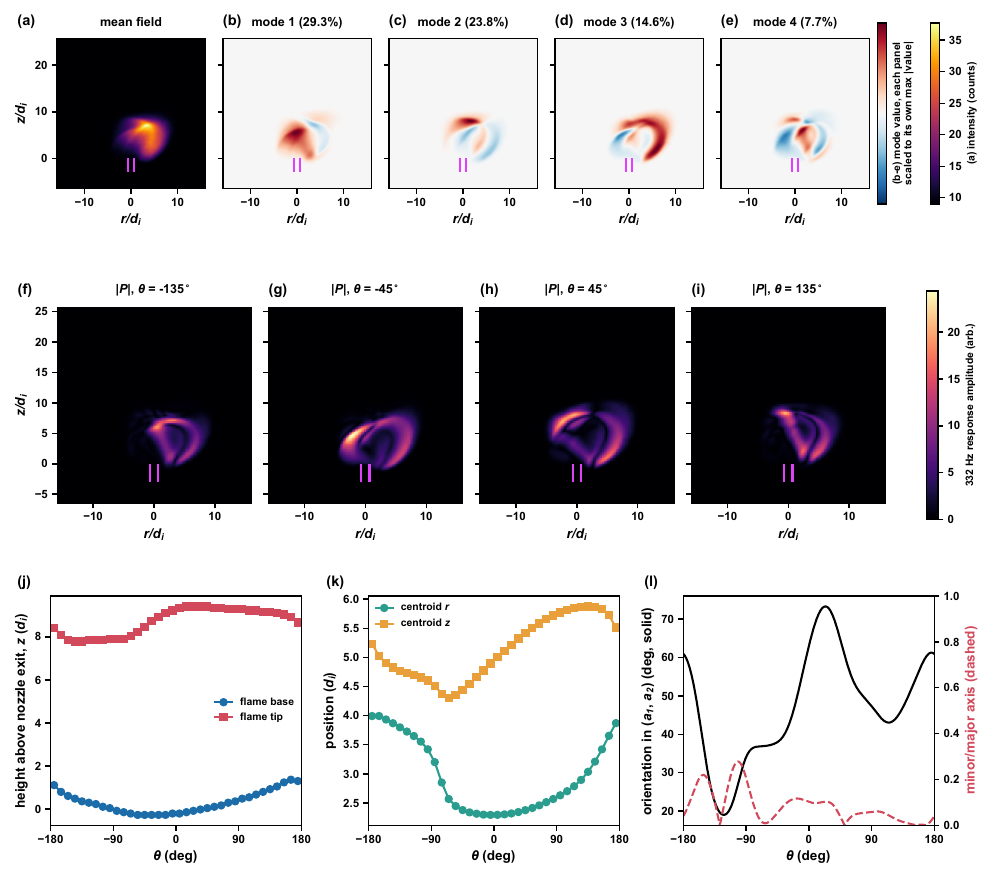}
\caption{Spatial structure of the PLOR cycle at $Re = 60$, $p' = 34.4$~Pa, in nozzle-centered coordinates (tube walls in magenta). (a)~Mean field; (b--e)~POD modes 1--4 with the fraction of the intensity variance they account for, each scaled to its own maximum; (f--i)~amplitude $|P|$ of the forced response~\eqref{eq:pattern} at four phases of the slow cycle; (j)~height of the flame base and tip and (k)~intensity centroid against the phase of the slow cycle; (l)~orientation (solid) and ratio of minor to major axis (dashed) of the forced oscillation in the plane of the two leading coefficients.}
\label{fig:physical}\end{figure}

\subsection{Evolution of the cycle with forcing amplitude}\label{sec:res-evolution}

At onset, the frequency of the slow cycle is nearly independent of the Reynolds number (15.9--16.7~Hz), but it decreases with increasing forcing amplitude, to 13.5, 12.7, and 10.4~Hz at the highest amplitudes for $Re$ = 60, 80, and 100, respectively (Figure~\ref{fig:onset}e; values for all conditions are listed in Table~\ref{tab:slow}, with uncertainties not exceeding 0.12~Hz). The decrease is not strictly monotonic: at $Re$ = 80 and 100, the frequency rises by 0.5 and 1.4~Hz, respectively, between two successive amplitudes, by more than its uncertainty. The decrease is not distributed uniformly over the cycle; it arises from the slow portion identified above, which becomes progressively slower and longer: at $Re = 100$, the ratio of the slowest to the mean rotation rate falls from 0.55 at onset to 0.09 at the highest amplitude, and the dwell fraction rises from 0.36 to 0.61 (Figures~\ref{fig:onset}f and \ref{fig:slowcycle}d). The same trend is observed at $Re$ = 60 and 80. A decrease of the low frequency with forcing amplitude, from about 16--17~Hz to about 14~Hz, was also reported for this burner at $Re = 65$ \cite{sim2020}. At the single PLOR condition for $Re = 40$, which immediately precedes blow-off, the cycle already has a pronounced slow portion ($\omega_{\min}/\langle\omega\rangle = 0.15$, dwell fraction 0.54).

The slow portion remains associated with the lifted flame as the forcing increases. Over all PLOR conditions, the phase of slowest rotation and the phase at which the flame base is highest differ by a median of 12$^\circ$, and by less than $30^\circ$ in 15 of the 17 conditions; the largest difference occurs at the onset condition for $Re = 80$ ($p' = 36.2$~Pa). The excursion of the flame base over the cycle ranges from 1.5~$d_i$ (at $Re = 40$) to 14.8~$d_i$ (at the highest amplitude for $Re = 100$) and generally increases with forcing amplitude at each Reynolds number. As the forcing increases, the flame is therefore lifted further and remains lifted for a growing fraction of a cycle whose period lengthens accordingly.

\subsection{Modulation of the forced response}\label{sec:res-forced}

\begin{figure}[tbp]\centering\includegraphics[width=\textwidth]{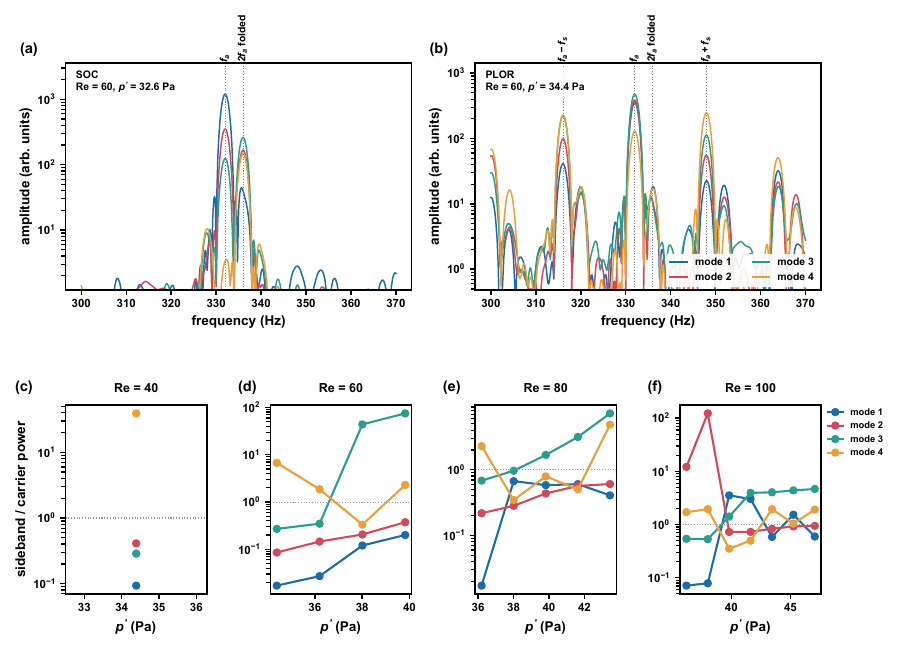}
\caption{Fast response. (a,~b)~Amplitude spectra of the fast parts of the four leading coefficients near the forcing frequency for SOC ($p' = 32.6$~Pa) and PLOR ($p' = 34.4$~Pa) at $Re = 60$; dotted lines mark $f_a$, the folded second harmonic $f_r - 2f_a$ and, in PLOR, $f_a \pm f_s$. (c--f)~Ratio of the power in the lines at $f_a \pm f_s$ to that in the line at $f_a$ for every PLOR condition at $Re$ = 40, 60, 80, and 100; colors identify POD modes 1--4, as in (a) and (b).}
\label{fig:fast}\end{figure}

In SOC the fast parts of the coefficients oscillate at the forcing frequency (Figure~\ref{fig:fast}a). The folded second harmonic, at $f_r - 2f_a$, is strong in the higher modes: in modes 3 and 4 at $Re = 60$, $p' = 32.6$~Pa, its amplitude is 2.1 and 42 times that of the forcing line, respectively. If the folding were not accounted for, the response of these modes would be attributed to the forcing frequency.

In PLOR the forced response is modulated by the slow cycle (Figure~\ref{fig:fast}b). In all four leading modes, lines appear at $f_a \pm f_s$ with comparable amplitudes above and below the forcing frequency, together with weaker lines at $f_a \pm 2f_s$. Where the modulation is strong, the forcing line itself is weak: in mode~4 at $Re = 60$, $p' = 34.4$~Pa, the power in the two sidebands is 6.7 times that of the forcing line. The mode in which the sidebands are relatively strongest changes with the Reynolds number and the amplitude (Figure~\ref{fig:fast}c--f). The sidebands therefore reflect a modulation of the forced response as a whole by the slow cycle rather than a property of any particular mode. The model~\eqref{eq:forced} accounts for at least 0.91 of the fast variance of modes 1--4 in every PLOR record and at least 0.89 on held-out data; the folded second harmonic contributes at most 3\% of the fast variance in PLOR.

The spatial pattern of the forced response~\eqref{eq:pattern} is concentrated along the flame edges and changes with the phase of the slow cycle (Figure~\ref{fig:physical}f--i). Accordingly, the forced oscillation in the plane of the two leading coefficients has no fixed orientation: at $Re = 60$, $p' = 34.4$~Pa, its orientation varies over a range of 54$^\circ$ during the slow cycle, about a mean of 47$^\circ$ (Figure~\ref{fig:physical}l).

\subsection{Spatial structure across Reynolds numbers}\label{sec:res-spatial}

\begin{figure}[tbp]\centering\includegraphics[width=\textwidth]{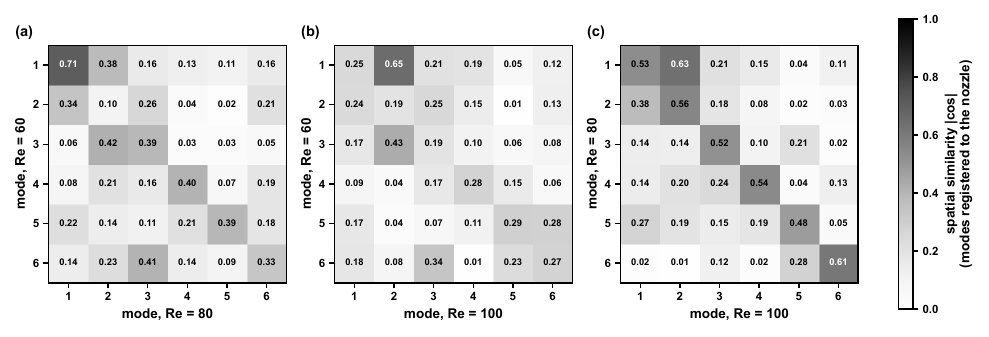}
\caption{Similarity of POD modes 1--6 between Reynolds numbers at the first PLOR condition, measured by the modulus of the normalized inner product after registration to nozzle-centered coordinates: (a)~$Re$ = 60 and 80, (b)~60 and 100, (c)~80 and 100.}
\label{fig:similarity}\end{figure}

Because the position of the nozzle in the image differs between the series recorded at different Reynolds numbers, the POD modes are compared after resampling onto common nozzle-centered coordinates (Figure~\ref{fig:similarity}). The leading mode at the first PLOR condition is similar between neighboring Reynolds numbers: the modulus of the normalized inner product is 0.86 between $Re$ = 40 and 60 (not shown in the figure), 0.71 between 60 and 80, and 0.53 between 80 and 100. The higher modes differ considerably more, and the ordering of the modes is not preserved. The distribution of the intensity variance among the modes therefore changes with Reynolds number; for this reason, the amplitude of the slow cycle is measured by the basis-invariant quantity~\eqref{eq:A} rather than by the coefficients of particular modes.

\subsection{Reduced description and its validation}\label{sec:res-validation}

\begin{figure}[tbp]\centering\includegraphics[width=\textwidth]{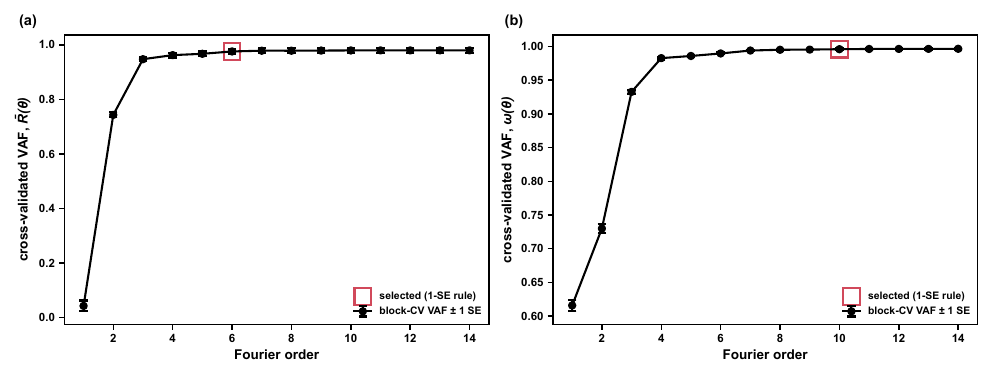}
\caption{Selection of the Fourier orders of (a)~$\bar R(\theta)$ and (b)~$\omega(\theta)$ by cross-validation over four contiguous blocks of the record ($Re = 60$, $p' = 34.4$~Pa). Error bars, one standard error; squares, selected orders.}
\label{fig:order}\end{figure}

\begin{figure}[tbp]\centering\includegraphics[width=\textwidth]{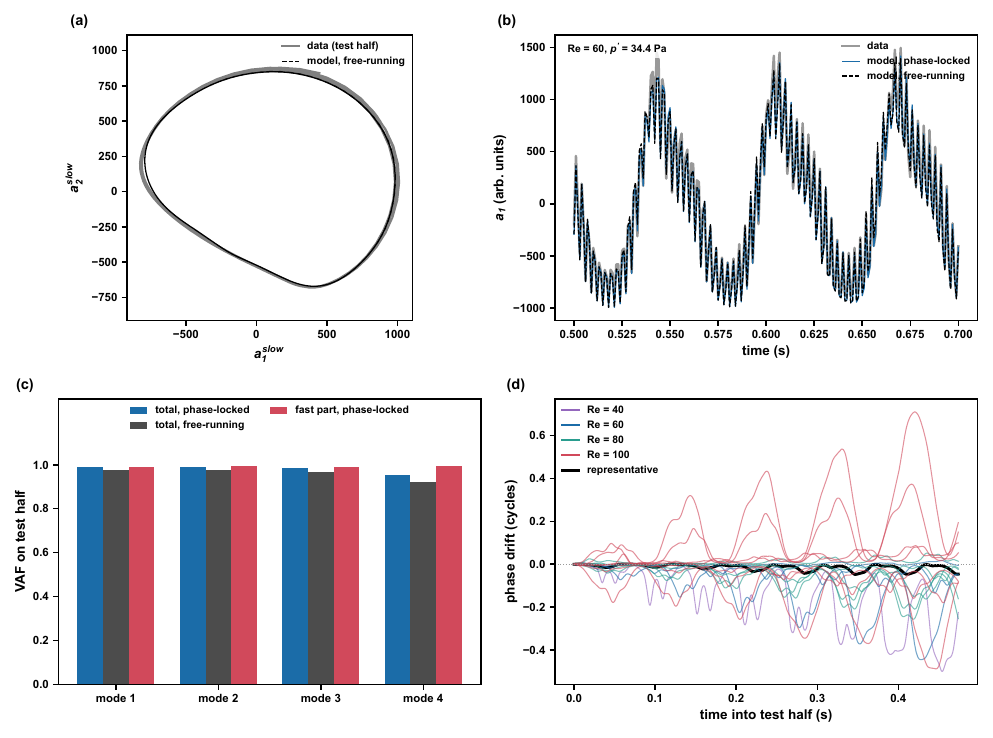}
\caption{Prediction of the held-out half of each record. (a)~Slow orbit and (b)~coefficient $a_1$, measured and predicted, at $Re = 60$, $p' = 34.4$~Pa; (c)~variance accounted for in modes 1--4 by the phase-locked and the free-running predictions; (d)~phase error of the free-running prediction for all PLOR conditions.}
\label{fig:validation}\end{figure}

Together, the slow cycle~\eqref{eq:fourier}, the slaved modes, and the modulated forced response~\eqref{eq:forced} constitute a reduced description of the four leading coefficients in which a single phase variable organizes both timescales. Its orders are selected by cross-validation (Figure~\ref{fig:order}), and its predictions on the held-out half of each record are shown in Figure~\ref{fig:validation}. When the observed phase is used, the description accounts for at least 0.90 of the variance of each of the four coefficients at every PLOR condition. When the phase is instead obtained by integrating $\omega(\theta)$, the description remains accurate near onset, accounting for at least 0.88 of the variance at the first PLOR condition for $Re$ = 60, 80, and 100, with phase errors of at most 0.25 cycles over the 5--8 cycles of the test interval. Its accuracy decreases at higher amplitudes, where the slow portion of the cycle is most pronounced: there, a small error in the minimum of $\omega(\theta)$ produces a large error in the period, and the frequency implied by the Fourier description exceeds the measured one by up to 0.65~Hz. The same applies to the single PLOR condition at $Re = 40$.

\section{Discussion}\label{sec:discussion}

\subsection{Nature of the onset}\label{sec:disc-onset}

Three observations bear on the nature of the SOC--PLOR transition. The slow amplitude changes by more than an order of magnitude within one amplitude step; the PLOR branch, continued below the onset, reaches zero below the last SOC condition; and the attached flame shows no growth of slow fluctuations as the onset is approached. Each is expected for a subcritical, hysteretic onset, and none is expected in the vicinity of a supercritical Hopf bifurcation, where the amplitude grows continuously from zero and the fluctuations of the stable state increase as its damping vanishes \cite{noiray2013}. Consistent with this, a square-root law with its critical amplitude inside the observed interval fits the PLOR branch less well than the unconstrained linear extrapolation (Section~\ref{sec:res-onset}).

The data nevertheless do not establish hysteresis. A supercritical branch that rose steeply within the last amplitude step of 1.8~Pa would be indistinguishable from a jump at this resolution, and a branch that levels off at high amplitude would also extrapolate below its true onset. Fluctuations of the attached flame are observable only if they are excited sufficiently by noise, and a loss of damping confined to the last amplitude step would escape detection in one-second records. Most importantly, bistability can be demonstrated only by approaching the same conditions from both directions: if the onset is hysteretic, PLOR should persist when the amplitude is decreased below the onset amplitude observed with increasing forcing. We therefore conclude that the onset is abrupt, is not preceded by a measurable change in the attached flame, and is consistent with a subcritical transition of the kind found in thermoacoustic systems with bistability \cite{subramanian2013,etikyala2017,junipersujith2018}.

The nature of the transition also differs from that of the forced self-excited flames and jets studied by \citet{lijuniper2013a,lijuniper2013b}, \citet{kashinath2018} and \citet{guan2019}. In those systems a natural oscillation exists without forcing, and increasing the forcing amplitude converts a quasi-periodic response into one locked to the forcing. Here the unforced flame has no slow oscillation; the forced flame first responds periodically, at the forcing frequency, and the slow oscillation appears only above a threshold, producing a quasi-periodic response. In dynamical-systems terms, the transition takes the periodic forced response to a quasi-periodic state, that is, to motion on a torus \cite{kuznetsov2004}, and the observations above characterize its onset. The dependence on Reynolds number is consistent with the stability maps of \citet{vargas2023}: at $Re = 20$ the flame blows off without entering PLOR, at $Re = 40$ PLOR is observed at a single condition, and at higher Reynolds numbers the PLOR range extends over several amplitude steps before blow-off.

\subsection{Interpretation of the slow phase}\label{sec:disc-lifted}

The central observation of this work is that the slow portion of the PLOR cycle coincides with the lifted state of the flame, and that this portion lengthens as the forcing increases while lift-off and reattachment remain comparatively rapid. The lengthening of the lifted phase accounts for the decrease of the PLOR frequency with forcing amplitude, which is consistent with the amplitude-dependent lift-off frequencies of 10--20~Hz reported for forcing near a pressure node \cite{hayrapetyan2025}. A frequency that decreases because the system spends increasingly long times in one part of its cycle is characteristic of oscillators approaching a bifurcation at which the period diverges \cite{strogatz2015}; in the present data, however, blow-off occurs while the frequency is still finite, so the data cannot show whether the period would diverge.

A physical interpretation consistent with these observations is that the lifted flame, whose base is stabilized away from the burner, can return to the burner rim only when local conditions---the forcing-induced flow and the mixture at the base---permit upstream propagation of its base. Laminar lifted flames are stabilized at a tribrachial edge whose position is set by a balance between propagation and the local flow \cite{chung1991}, and forced lifted flames can alternate between attached and lifted states \cite{baillot2002,baillot2007}; stronger forcing would be expected both to lift the flame further and to delay its return. The present observations do not identify the mechanism; doing so would require velocity and mixture-fraction measurements at the flame base. The behavior contrasts with that near a pressure antinode, where the perturbations are symmetric and PLOR appears predominantly as a period doubling at half the forcing frequency \cite{hayrapetyan2025}.

\subsection{Scope of the reduced description}\label{sec:disc-scope}

The phase-resolved description summarizes the PLOR attractor at each condition: the shape of the slow cycle, the rate at which it is traversed, the dependence of the higher modes on its phase, and the modulation of the forced response. It predicts held-out data from the same record, and its timing is accurate near onset; it is not, however, a model of the transition. Data recorded on an attractor determine the cycle but not the dynamics away from it, which govern its stability and its emergence; a normal form fitted to such data, such as a cubic Stuart--Landau equation, cannot be identified uniquely, and the cubic form would in any case be unable to represent a subcritical onset, which requires saturation at higher order \cite{subramanian2013}. A predictive model of the transition would require information about the response to perturbations---transients after changes in forcing amplitude, or measurements taken while the amplitude is decreased---together with parameter-dependent fits across conditions. The same limitation applies to sparse-regression models of the PLOR regime \cite{brunton2016,lettieri2025}.

\subsection{Implications for the analysis of similar data}\label{sec:disc-methods}

Several features of the data required specific treatment and are likely to arise in other analyses of high-speed imaging of forced flames. When the frame rate is about three times the forcing frequency, harmonics of the forced response fold into the measured band; here the second harmonic appears next to the forcing frequency and the third inside the slow band, where it can be mistaken for a precursor of the slow oscillation. Phase-based measures of the shape of a limit cycle depend on the reference point; the POD origin, being the time average, is displaced toward the slowly traversed parts of the orbit. In short records, the end effects of zero-phase filters scale with the amplitude of the forced response and can create spurious trends in quantities computed from the slow band, which windowed spectral estimates avoid. Finally, comparisons of spatial modes between separate runs require registration of the images, since the position of the nozzle in the frame may change between runs.

\subsection{Limitations}\label{sec:disc-limits}

The analysis is limited by the available data. The forcing amplitude was only increased, so hysteresis cannot be tested directly. The amplitude steps of 1.8~Pa set the resolution with which the onset can be located. A single position of the injector relative to the pressure node was examined. The local pressure amplitude is inferred from the maximum amplitude in the waveguide using the calibration of the original experiments. Spatial quantities are computed from 20-mode reconstructions and rely on a calibration of the image coordinates with a positional uncertainty of about 0.2~$d_i$.

\section{Conclusions}\label{sec:conclusions}

High-speed imaging of a methane microjet flame forced near the pressure node of a 332~Hz standing wave has been analyzed over the full range of forcing amplitudes at five Reynolds numbers. SOC and PLOR are sharply distinct, with slow-variance fractions of at most 0.3\% and at least 67\%, respectively. PLOR appears abruptly: the slow amplitude increases by a factor of 14--37 within one amplitude step, the attached flame shows no growth of slow fluctuations beforehand, and the PLOR branch extrapolates to zero below the observed onset. These observations are consistent with a hysteretic onset rather than with a supercritical Hopf bifurcation; experiments in which the forcing amplitude is also decreased would be required to confirm hysteresis.

The PLOR cycle is organized by a single phase variable, and the phase at which it is traversed most slowly coincides with the greatest lift-off of the flame base. As the forcing increases, the excursion of the flame base grows, reaching 14.8 nozzle diameters, and the flame remains lifted for a growing fraction of the cycle, spending up to 61\% of the period within one quarter of the cycle; the PLOR frequency correspondingly decreases from about 16~Hz at onset to 10.4~Hz. The forced response is modulated in amplitude and phase by the slow cycle, producing sidebands at $f_a \pm f_s$ in all leading modes. A phase-resolved description built on these observations reproduces held-out data with a variance accounted for of at least 0.90 when phase-locked, and its free-running timing is accurate near onset but degrades as the slow lifted phase becomes more pronounced.

\section*{Acknowledgments}
The author thanks A.~Vargas and A.~R.~Karagozian for providing the data analyzed in this work.

\section*{Data availability}
The analysis code and the derived results are available from the author on reasonable request.

\bibliographystyle{unsrtnat}
\bibliography{references}

\appendix
\renewcommand{\thefigure}{A\arabic{figure}}\setcounter{figure}{0}
\renewcommand{\thetable}{A\arabic{table}}\setcounter{table}{0}
\section{Supplementary material}\label{app:supp}

\begin{table}[h]\centering\small
\caption{Slow cycle for every PLOR condition: frequency with its block standard error; ratio of the slowest to the mean rotation rate and dwell fraction, both from the phase density~\eqref{eq:density}; phase of slowest rotation, $\theta_{\mathrm{slow}}$; phase at which the flame base is highest, $\theta_{\mathrm{lift}}$; and excursion of the flame base over the cycle.}
\label{tab:slow}
\begin{tabular}{rrcccccc}
\toprule
$Re$ & $p'$ & $f_s$ & $\omega_{\min}/\langle\omega\rangle$ & dwell & $\theta_{\mathrm{slow}}$ & $\theta_{\mathrm{lift}}$ & base excursion \\
 & (Pa) & (Hz) & & & ($^\circ$) & ($^\circ$) & ($d_i$) \\
\midrule
40 & 34.4 & 16.69$\pm$0.07 & 0.15 & 0.54 & $-$168 & $-$175 & 1.5 \\
60 & 34.4 & 15.94$\pm$0.04 & 0.40 & 0.36 & 178 & 165 & 1.6 \\
60 & 36.2 & 16.11$\pm$0.11 & 0.29 & 0.42 & 178 & 165 & 2.8 \\
60 & 38.0 & 14.38$\pm$0.02 & 0.20 & 0.49 & $-$172 & 175 & 5.2 \\
60 & 39.8 & 13.45$\pm$0.09 & 0.17 & 0.53 & $-$172 & 175 & 7.2 \\
80 & 36.2 & 16.18$\pm$0.04 & 0.47 & 0.33 & $-$108 & 135 & 2.3 \\
80 & 38.0 & 15.64$\pm$0.06 & 0.49 & 0.37 & 178 & 165 & 4.3 \\
80 & 39.8 & 13.22$\pm$0.05 & 0.25 & 0.48 & $-$178 & 165 & 7.4 \\
80 & 41.6 & 13.74$\pm$0.04 & 0.27 & 0.48 & $-$178 & 165 & 7.5 \\
80 & 43.4 & 12.68$\pm$0.04 & 0.16 & 0.55 & 178 & 165 & 9.4 \\
100 & 36.2 & 16.07$\pm$0.09 & 0.55 & 0.36 & $-$128 & $-$95 & 3.2 \\
100 & 38.0 & 15.97$\pm$0.09 & 0.59 & 0.34 & $-$123 & $-$95 & 3.2 \\
100 & 39.8 & 15.34$\pm$0.09 & 0.51 & 0.37 & 172 & 155 & 5.8 \\
100 & 41.6 & 13.11$\pm$0.08 & 0.28 & 0.48 & $-$178 & 165 & 9.7 \\
100 & 43.4 & 11.11$\pm$0.11 & 0.12 & 0.58 & $-$172 & $-$165 & 14.0 \\
100 & 45.2 & 12.54$\pm$0.11 & 0.17 & 0.52 & 178 & 165 & 11.9 \\
100 & 47.1 & 10.42$\pm$0.12 & 0.09 & 0.61 & $-$172 & $-$165 & 14.8 \\
\bottomrule
\end{tabular}\end{table}

\begin{figure}[h]\centering\includegraphics[width=\textwidth]{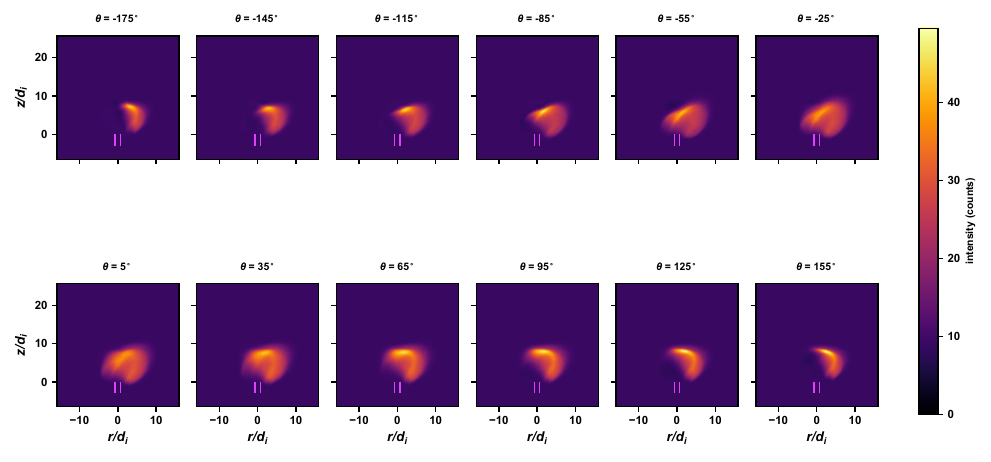}
\caption{Phase-averaged reconstructions of the flame (mean field and 20 modes) at twelve phases of the slow cycle, $Re = 60$, $p' = 34.4$~Pa, in nozzle-centered coordinates (tube walls in magenta).}
\label{fig:S1}\end{figure}

\begin{figure}[h]\centering\includegraphics[width=\textwidth]{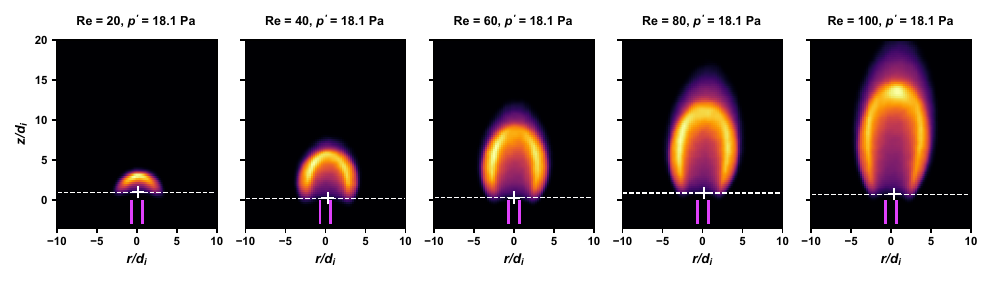}
\caption{Mean fields at the lowest forcing amplitude ($p' = 18.1$~Pa) for each Reynolds number in nozzle-centered coordinates; dashed lines, detected height of the flame base; crosses, intensity-weighted lateral centroid of the flame, plotted at the height of its base.}
\label{fig:coords}\end{figure}

\end{document}